\documentstyle[12pt,amsfonts]{article}
\def\one{1\hskip-.37em 1}

\def\half{\textstyle{\frac{1}{2}}}
\def\quarter{\textstyle{\frac{1}{4}}}

\def\H{{\cal H}}
\def\l{\lambda}
\def\p{\phi}
\def\D{{\cal D}}
\def\g{\gamma}
\def\s{\hskip.013cm}

\def\E{{\rm I}\hskip-.2em{\rm E}}
\def\ra{\rightarrow}
\def\tint{{\textstyle\int}}
\def\dd{\partial}
\def\o{\overline}
\def\a{\alpha}

\def\b{\begin{eqnarray*}}     
\def\e{\end{eqnarray*}}       
\def\bn{\begin{eqnarray}}     
\def\en{\end{eqnarray}}       
\def\<{\langle}
\def\>{\rangle}

\def\no{\nonumber}

\def\{{\lbrace}
\def\}{\rbrace}
\def\hg{{\hat g}}
\def\hp{{\hat\pi}}

\begin{document}

\date{}

\title{Fundamentals of Quantum Gravity\footnote{Presented at the Workshop ``Particles and Fields: Classical and Quantum'',
September, 2006, in Jaca, Spain, in honor of the 75th birthday of
E.C.G. Sudarshan.}}
\author{John R. Klauder\footnote{Electronic mail: klauder@phys.ufl.edu}\\
Department of Physics and Department of Mathematics\\
University of Florida\\
Gainesville, FL  32611}

\maketitle

\begin{abstract}
The outline of a recent approach to quantum gravity is presented.
Novel ingredients include: (1) Affine kinematical variables; (2)
Affine coherent states; (3) Projection operator approach toward
quantum constraints; (4) Continuous-time regularized functional
integral representation without/with constraints; and (5) Hard core
picture of nonrenormalizability. The ``diagonal representation'' for
operator representations, introduced by Sudarshan into quantum
optics, arises naturally within this program.

\end{abstract}


\section*{Introduction}
Nearly 40 years ago, George Sudarshan and the present author
published our book ``Fundamentals of Quantum Optics'', \cite{book1}
and it is noteworthy that this book has been recently
reprinted by Dover \cite{book1a}.
 The title of the present paper is
meant to honor the title of our earlier joint work, but in fact, it
is also meant in a literal sense as well in that the approach to be
outlined in this paper does constitute, in the author's opinion,  a
fundamental view of how quantum gravity can be approached. It is
important at the outset to remark that what is presented here is not
string theory nor is it loop quantum gravity, the two most commonly
studied approaches to quantum gravity. In the author's judgment,
the present approach, known as Affine Quantum Gravity, is more
natural than either of the traditional views and is closer to
classical (Einstein) gravity as well. General references for
this paper are \cite{kla1,kla2,kla3}.

The paper is divided into several sections each representing a
fundamental building block in the edifice we hope to construct. The
basic building blocks are designed to address essential components
of any natural approach to a quantum field theory. Section 1
addresses the question of just what constitutes the proper choice of
a fundamental set of kinematical, phase space, field variables.

In Sec.~2 we first observe that the gravitational field theory is
a special theory in that all its dynamical content is enforced by
constraints. Since quantization normally requires a phase space
geometry with a high degree of symmetry, it follows that it is
generally prudent to quantize first and reduce second because if one
reduces first, there is generally no guarantee that the reduced
classical phase space still has sufficient symmetry to ensure an
ambiguity-free quantization. Consequently, in Sec.~2 we choose a
representation of the field operators (from among uncountably many
inequivalent, irreducible choices!) before we have imposed any of
the constraints.

Section 3 takes up the question of the quantum constraints. It is
characteristic of gravity that it is classically
 an open first-class constraint system, meaning that it fulfills a
 Lie-algebra like set of mutual Poisson brackets among the
 constraints save for the fact that instead of structure constants
 there are structure functions of the phase space variables. When
 quantized, these structure functions become operators which, in
 the case of gravity, do not commute with the constraints and
 thereby lead to a set of quantum constraints that are partially
 second class in nature. Normally, such an anomalous behavior
 requires the introduction of unphysical, auxiliary variables,
 if it can be treated at all. However, there is a relatively new
 approach to deal with operator constraints that treats both first-
 and second-class constraints in exactly the same way, and as a
 consequence, this method, which is explained in Sec.~3, seems to
 be ideal to analyze the gravitational field.

 Functional integral methods are valuable as guides in quantization.
 Sometimes, one is fortunate to actually evaluate the integral, but
 even when that is not possible, the form of the integral itself can
 sometimes be used to draw useful qualitative conclusions. Therefore,
 it is important to observe that the present formulation of quantum
 gravity admits a reasonably well-defined functional integral, both
in the initial case without imposition of any constraints, as well as
in the case in which the constraints are introduced; this general
subject is discussed in Sec.~4.

Finally, we face the conventional wisdom that gravity is a
perturbatively nonrenormalizable theory. To deal with this situation
we recall the hard core theory of nonrenormalizable theories in
general. This theory asserts that nonrenormalizable quantum field
theories behave as they do because, from a functional integral point
of view, the nonlinear interaction term acts as a hard core
projecting out field distributions that would otherwise have been allowed by
the free theory alone. As a consequence, as with any hard-core
interaction, the interacting theory does {\it not} reduce to the
free theory (in the sense of Green's function convergence) as the
coupling constant vanishes, and thus
the use of regularized perturbation about the free theory to suggest
counterterms to the quantum theory is inappropriate. In Sec.~5, we outline
the hard-core theory for quantum gravity, which although we do not
have explicit control of such hard cores is nonetheless highly
suggestive.

\section{Affine Kinematical Variables}

\subsubsection*{Metric positivity}
An essential property of affine quantum gravity is the strict
positivity of the spatial metric. For the classical metric, this
property means that for any nonvanishing set $\{u^a\}$ of real
numbers and any nonvanishing, nonnegative test function, $f(x)\ge0$,
that
 \bn  \tint f(x)\s u^a g_{ab}(x)\s u^b\s d^3\!x>0\;,  \en
where $1\le a,b \le3$.
We  also insist that this inequality holds when the classical metric field
$g_{ab}(x)$ is replaced with the
$3\times3$ operator metric field $\hg_{ab}(x)$.

\subsubsection*{Affine commutation relations}
The canonical commutation relations are not compatible with the
requirement of metric positivity since the canonical momentum acts
to translate the spectrum of the metric tensor, and such a
translation is incompatible with metric positivity. Thus it is
necessary to find a suitable but distinctly alternative set of
commutation relations. A suitable alternative that has the virtue of
preserving the spectrum of a positive definite metric operator is readily
available.

The initial step involves replacing the classical ADM canonical
momentum $\pi^{ab}(x)$ with the classical mixed-index momentum
$\pi^a_b(x)\equiv \pi^{ac}(x)g_{cb}(x)$. We refer to $\pi^a_b(x)$ as
the ``momentric" tensor being a combination of the canonical {\it
momen}tum and the canonical me{\it tric}.
Besides the metric being promoted to an operator $\hg_{ab}(x)$, we
also promote the classical momentric tensor to an operator field
$\hp^a_b(x)$; this pair of operators form the basic kinematical
affine operator fields, and all operators of interest are given as
functions of this fundamental pair. The basic kinematical operators
are chosen so that they satisfy the following set of {\it affine
commutation relations} (in units where $\hbar=1$, which are normally
 used throughout): \bn
  &&\hskip.191cm[\hp^a_b(x),\,\hp^c_d(y)]=\half\s i\s[\s\delta^c_b\hp^a_d(x)-
  \delta^a_d\hp^c_b(x)\s]\,\delta(x,y)\;,\no\\
  &&\hskip.1cm[\hg_{ab}(x),\,\hp^c_d(y)]=\half\s i\s[\s\delta^c_a\hg_{bd}(x)+
  \delta^c_b\hg_{ad}(x)\s]\,\delta(x,y)\;,\label{afc}\\
&&[\hg_{ab}(x),\,\hg_{cd}(y)]=0\;.   \no \en These commutation
relations arise as the transcription into operators
of equivalent Poisson brackets for the
corresponding classical fields, namely, the spatial metric
$g_{ab}(x)$ and the momentric field
$\pi^c_d(x)\equiv \pi^{cb}(x)\s g_{bd}(x)$, along with the usual
Poisson brackets between the canonical metric field $g_{ab}(x)$ and the
canonical momentum field $\pi^{cd}(x)$.

The virtue of the affine variables and their associated commutation
relations is evident in the relation
\bn e^{ i\tint\gamma^a_b(y)\s\hp^b_a(y)\,d^3\!y}\,\hg_{cd}(x)\,
  e^{-i\tint\gamma^a_b(y)\s\hp^b_a(y)\,d^3\!y}=
\{\s e^{\g(x)/2}\s\}_c^e\,\hg_{ef}(x)\,\{\s e^{\g(x)^T/2}\s\}_d^f\;, \en
where $\g^T(x)$ denotes the transpose of the matrix $\g(x)$. This
algebraic relation
confirms that suitable transformations by the momentric field preserve metric
positivity.

\section{Affine Coherent States}
It is noteworthy that the algebra generated by $\hg_{ab}$ and
$\hp^a_b$ as represented by (\ref{afc}) closes. These operators form
the generators of the {\it affine group} whose elements may be
defined by
  \bn U[\pi,\gamma]\equiv e^{i\tint \pi^{ab}(y)\s\hg_{ab}(y)\,d^3\!y}\,
  e^{-i\tint\gamma^a_b(y)\s\hp^b_a(y)\,d^3\!y}\;,\en
e.g., for all real, smooth $c$-number functions $\pi^{ab}$ and
$\gamma^a_b$ of compact support. Since we assume that the smeared
fields $\hg_{ab}$ and $\hp^a_b$ are self-adjoint operators, it
follows that $U[\pi,\gamma]$ are unitary operators for all $\pi$ and
$\gamma$, and moreover, these unitary operators are strongly continuous
in the label fields $\pi$ and
$\gamma$.

To define a representation of the basic operators it suffices to choose
a fiducial vector and thereby to introduce a set of affine
coherent states, i.e., coherent states formed with the help of the
affine group. We choose $|\s\eta\>$ as a normalized fiducial vector
in the original Hilbert space $\frak H$, and we consider a
set of unit vectors each of
which is given by \bn
  |\pi,\gamma\>\equiv e^{i\tint \pi^{ab}(x)\,\hg_{ab}(x)\,d^3\!x}\,e^{-i\tint\gamma^d_c(x)
  \,{\hat\pi}^c_d(x)\,d^3\!x}\,|\s\eta\>\;.  \en
As $\pi$ and $\gamma$ range over the space of smooth
functions of compact support, such vectors form the desired set of
coherent states. The specific representation of the kinematical
operators is fixed once the vector $|\s\eta\>$ has been chosen.
As minimum requirements on $|\s\eta\>$ we impose
         \bn &&\hskip.041cm\<\eta|\hp^a_b(x)|\s\eta\>=0\;, \\
             &&\<\eta|\hg_{ab}(x)|\eta\>={\tilde g}_{ab}(x)\label{t8}\;, \en
where ${\tilde g}_{ab}(x)$ is a metric that determines the topology
of the underlying space-like surface.
As algebraic consequences of these conditions, it follows that
  \bn
   && \hskip.07cm\<\pi,\g|\s \hg_{ab}(x)\s|\pi,\g\>=\{\s e^{\g(x)/2}\s\}_a^c\,{\tilde g}_{cd}(x)\,\{\s e^{\g(x)^T/2}\s\}_b^d\equiv g_{ab}(x)\label{t34}\;,  \\
   &&\hskip.14cm\<\pi,\g|\s{\hat\pi}^a_c(x)\s|\pi,\g\>=\pi^{ab}(x)\s g_{bc}(x)\equiv \pi^a_c(x)\label{t35}\;. \en
 These expectations are not gauge
invariant, nor should they be, since they are taken in the
original Hilbert space where the constraints are not fulfilled.

By definition, the coherent states span the original, or kinematical,
Hilbert space $\frak H$, and thus we can characterize the coherent
states themselves by giving their overlap with an arbitrary coherent
state. In so doing, we choose the fiducial vector $|\s\eta\>$ so that the
overlap is given  by
 \bn &&\hskip-.3cm\<\pi'',\gamma''|\pi',\gamma'\>
   =\exp\bigg[-2\int b(x)\,d^3\!x\,\no\\
&&\hskip.1cm\times\ln\bigg(\frac{\det\{\half[g''^{ab}(x)+g'^{ab}(x)]+\half
ib(x)^{-1}
[\pi''^{ab}(x)-\pi'^{ab}(x)]\}}{\{\det[g''^{ab}(x)]\,\det[g'^{ab}(x)]\}^{1/2}}\bigg)
\bigg]\;, \label{t18}\en
where $b(x)$, $0<b(x)<\infty$, is a scalar density which is discussed below.

 Additionally, we
observe that $\gamma''$ and
$\gamma'$ do {\it not} appear in the explicit functional form
given in (\ref{t18}).
In particular, the smooth matrix $\gamma$ has been replaced by the
smooth matrix $g$ which is defined at every point by
 \bn  g(x)\equiv e^{\gamma(x)/2}\,{\tilde g}(x)\,e^{\gamma(x)^T/2}\equiv\{g_{ab}(x)\}\;, \en
where the matrix ${\tilde g}(x)\equiv \{{\tilde g}_{ab}(x)\}$ is
given by (\ref{t8}).
 The map $\gamma\ra g$ is clearly
many-to-one since $\gamma$
 has nine independent variables at each point while $g$, which is symmetric, has only six.
In view of this functional dependence we may denote the given
functional in (\ref{t18}) by $\<\pi'',g''|\pi',g'\>$, and henceforth
we shall adopt this notation. In particular, we note
that (\ref{t34}) and (\ref{t35}) become
   \bn &&
\hskip.07cm\<\pi,g|\s \hg_{ab}(x)\s|\pi,g\>\equiv g_{ab}(x)\;,  \\
   &&\hskip.143cm\<\pi,g|\s{\hat\pi}^a_c(x)\s|\pi,g\>=\pi^{ab}(x)
\s g_{bc}(x)\equiv \pi^a_c(x)\;, \en
which show that the meaning of the labels $\pi$ and $g$ is that
of {\it mean} values rather than sharp eigenvalues.

\subsubsection*{Reproducing kernel Hilbert spaces}
Although not commonly used, reproducing kernel Hilbert spaces are
very natural and readily understood. By definition, the vectors
$\{|\pi,g\>\}$ span the Hilbert space $\frak H$, and therefore
two elements of a
dense set of vectors have the form
   \bn
        &&|\phi\>=\sum_{j=1}^J\a_j\,|\pi_{[j]},g_{[j]}\>\;,\\
        &&|\psi\>=\sum_{k=1}^K\beta_k\,|\pi_{(k)},g_{(k)}\>\;,\en
for general sets  $\{\a_j\}_{j=1}^J$, $\{\beta_k\}_{k=1}^K$,
 $\{\pi_{[j]},g_{[j]}\}_{j=1}^J$, $\{\pi_{(k)},g_{(k)}\}_{k=1}^K$,
and some $J,K<\infty$. The inner
product of two such vectors is clearly given by
  \bn  \<\phi|\psi\>=\sum_{j,k=1}^{J,K}\s\a^*_j\beta_k\s\<\pi_{[j]},g_{[j]}|\pi_{(k)},g_{(k)}\>\;.
  \label{e6} \en

To {\it represent} the abstract vectors themselves as functionals,
we adopt the natural coherent-state representation, i.e.,
\bn&&\hskip.02cm \phi(\pi,g)\equiv \<\pi,g|\phi\>=\sum_{j=1}^J\a_j\,\<\pi,g|\pi_{[j]},g_{[j]}\>\;,\no\\
    &&\psi(\pi,g)\equiv \<\pi,g|\psi\>=\sum_{k=1}^K\beta_k\,\<\pi,g|\pi_{(k)},g_{(k)}\>\,.  \en
Thus, we have a dense set of continuous functions
 and a definition of an inner product
between pairs of such functions given by
     \bn (\phi,\psi)\equiv \<\phi|\psi\>\;,   \en
as defined in (\ref{e6}). It only remains to complete the space to a
(separable) Hilbert space $\frak C$, composed entirely of
continuous functions, by adding the limit points of all
Cauchy sequences in the norm $\|\psi\|\equiv(\psi,\psi)^{1/2}$.
 Note well that {\it all properties} of the
reproducing kernel Hilbert space $\frak C$ follow as direct
consequences from the continuous coherent-state overlap function
$\<\pi'',g''|\pi',g'\>$ itself; for details see, e.g., \cite{mesh}.

\section{Projection Operator Approach Toward \\Quantum Constraints}
Consider a classical phase space system with a set of constraints given by
$\phi_\a(p,q)=0$ for all $\a$, $1\le\a\le A$, which defines the
constraint hypersurface ${\cal C}\equiv\{(p,q):\phi_\a(p,q)=0,\,
 {\rm for\;all\;\a}\}$. Such constraints
are added to the classical Hamiltonian $H(p,q)$ with the help of
Lagrange multipliers $\{\l^\a(t)\}$ to form the total
Hamiltonian
   \bn  H_T(p,q)=H(p,q)+\l^\a\s\phi_\a(p,q)\;.  \en
The time derivative of the constraints
must vanish as well, and this condition leads to
\bn {\dot\phi}_\a=\{\phi_\a,\s H\s\}+\l^\beta\s\{\phi_\a,
\s\phi_\beta\}=0\;.\en
First class constraints arise when both Poisson brackets vanish on $\cal C$,
 and therefore
   \bn  &&  \{\phi_\a,\s\phi_\beta\}=c_{\a\s
\beta}^{\;\;\;\;\;\g}\s\phi_\g\;, \\
   && \hskip.066cm\{\phi_\a,\s H\s\}=h_\a^{\;\;\beta}\s\phi_\beta \;. \en
If $c_{\a\s\beta}^{\;\;\;\;\;\g}$ are constants, then the system
is called
closed first class; if instead $c_{\a\s\beta}^{\;\;\;\;\;\g}$ are
functions of the phase space variables, then the system is called
open first class. In either case, the Lagrange multipliers are not determined by the equations of motion and must be chosen (a ``gauge'' choice) to find the solution of the equations of motion.

Instead, if the Poisson bracket of the constraints does
{\it not} vanish on $\cal C$, assuming for illustration that it
has an inverse, then it follows that
  \bn \l^\beta\equiv -\s\{\phi_\a,\s\phi_\beta\}^{-1}\s\{\phi_\a,\s H\s\}\;,\en
which means that the Lagrange multipliers are determined by the equations of motion. In this case, the constraints are referred to as second class. Of course, there can be intermediate cases for which some of the constraints are first class while the remainder are second class.

The Dirac approach to the quantization of constraints requires quantization before reduction. Thus the constraints are first promoted to self-adjoint
operators,
  \bn \phi_\a(p,q)\ra\Phi_\a(P,Q)\;, \en
for all $\a$, and then the physical Hilbert space ${\frak H}_{phys}$ is defined by those vectors $|\psi\>_{phys}$  for which
   \bn \Phi_\a(P,Q)\s|\psi\>_{phys}=0\; \label{t77}\en
for all $\a$.
This procedure works for a limited set of classical first class constraint systems, but it does not work in general and especially not for second class constraints.

The projection operator approach to quantum constraints proceeds by offering a slight generalization of the Dirac procedure. Instead of insisting that
(\ref{t77}) holds exactly, we introduce a projection operator $\E$
defined by
  \bn  \E=\E(\Sigma_\a\Phi_\a^2\le\delta(\hbar)^2)\;,  \en
where $\delta(\hbar)$ is a positive {\it regularization parameter}
and we have assumed that
$\Sigma_\a\Phi_\a^2$ is self adjoint. This relation means that $\E$
projects onto the spectral range of the self-adjoint operator
$\Sigma_\a\Phi_\a^2$ in the interval $[0,\delta(\hbar)^2]$.
In this case, ${\frak H}_{phys}=\E\s{\frak H}$.
As a final step, the parameter $\delta(\hbar)$ is reduced as much as
required, and, in particular, when some second-class constraints are
involved, $\delta(\hbar)$ ultimately remains strictly positive. This
general procedure treats all constraints simultaneously and treats
them all on an equal basis; see \cite{sch}.

Several examples illustrate how the projection operator method works.
If $\Sigma_\a\s\Phi_\a^2=J_1^2+J_2^2+J_3^2$, the Casimir operator of
$su(2)$, then $0\le\delta(\hbar)^2<3\hbar^2/4$ works for this first
class example. If $\Sigma_\a\s\Phi_\a^2=P^2+Q^2$, where
$[Q,P]=i\hbar\one$, then $\hbar\le\delta(\hbar)^2<3\hbar$ covers
this second class example. If the single constraint $\Phi=Q$,
an operator whose zero lies in the continuous spectrum, then
it is convenient to take an appropriate form limit
of the projection operator as $\delta\ra0$; see \cite{sch}. The projection operator scheme can also deal with irregular constraints such as $\Phi=Q^3$, and even mixed examples with regular and irregular constraints such as
$\Phi=Q^3(1-Q)$, etc.; see \cite{kl-lit}.

It is also of interest that the desired projection operator
has a general, time-ordered integral representation (see \cite{kla6})
given by
  \bn  \E=\E(\!\!(\Sigma_\a\s\Phi_\a^2\le\s\delta(\hbar)^2\s)\!\!) =\int {\sf T}\s
  e^{-i\tint\lambda^\a(t)\s\Phi_\a\,dt}\,\D R(\lambda)\;. \label{e10}\en
The weak measure $R$ depends on the number of Lagrange multipliers,
the time interval, and the regularization parameter
$\delta(\hbar)^2$. The measure $R$ does {\it not} depend on the constraint operators, and thus this relation is an operator identity, holding for any set
of operators $\{\Phi_\a\}$.
The time-ordered integral representation for
$\E$ given in (\ref{e10}) can be used in path-integral representations as will
become clear below.

\section{Continuous-time Regularized Functional \\Integral Representation
without/with \\Constraints}
It is useful to reexpress the coherent-state overlap function by means
of a functional integral. This process can be aided by the fact that
the expression (\ref{t18}) is analytic in the variable $g''^{ab}(x)
+i\s b(x)^{-1}\s\pi''^{ab}(x)$ up to a factor. As a consequence the elements of the reproducing kernel Hilbert space satisfy a complex polarization condition, which leads to a second-order differential operator that annihilates each element of
$\frak C$. This fact can be used to generate a functional
representation of the form
  \bn  &&\<\pi'',g''|\pi',g'\>
              =\exp\bigg[-2\int b(x)\,d^3\!x\,\no\\
&&\hskip1.2cm\times\ln\bigg(\frac{\det\{\half[g''^{ab}(x)+g'^{ab}(x)]+\half
ib(x)^{-1}
[\pi''^{ab}(x)-\pi'^{ab}(x)]\}}{\{\det[g''^{ab}(x)]\,\det[g'^{ab}(x)]\}^{1/2}}\bigg)\bigg] \no\\
&&\hskip.8cm=\lim_{\nu\ra\infty}\,{\o{\cal N}}_{\nu}\,\int \exp[-i\tint g_{ab}
\s{\dot\pi}^{ab}\,d^3\!x\,dt]\no\\
  &&\hskip1.4cm\times\exp\{-(1/2\nu)\tint[b(x)^{-1}g_{ab}g_{cd}{\dot\pi}^{bc}{\dot\pi}^{da}+
  b(x)g^{ab}g^{cd}{\dot g}_{bc}{\dot g}_{da}]\,d^3\!x\,dt\}\no\\
&&\hskip2.3cm\times[\s \Pi_{x,t}\,\Pi_{a\le
b}\,d\pi^{ab}(x,t)\,dg_{ab}(x,t)\s] \label{e20}\;.  \en Here, because
of the way the new independent variable $t$ appears in the
right-hand term of this equation, it is natural to interpret $t$,
$0\le t\le T$, $T>0$ as coordinate ``time''. The fields on the
right-hand side all depend on space and time, i.e.,
$g_{ab}=g_{ab}(x,t)$, ${\dot g}_{ab}=\dd g_{ab}(x,t)/\dd t$, etc.,
and, importantly, the integration domain of the formal measure is
strictly limited to the domain where $\{g_{ab}(x,t)\}$ is a
positive-definite matrix for all $x$ and $t$. For the boundary
conditions, we have $\pi'^{ab}(x)\equiv\pi^{ab}(x,0)$,
$g'_{ab}(x)\equiv g_{ab}(x,0)$, as well as
$\pi''^{ab}(x)\equiv\pi^{ab}(x,T)$, $g''_{ab}(x)\equiv g_{ab}(x,T)$
for all $x$. Observe that the right-hand term holds for any
$T$, $0<T<\infty$, while the left-hand and middle terms are
independent of $T$ altogether.

In like manner, we can incorporate the constraints into a functional integral by using an appropriate form of the integral representation (\ref{e10}). The resultant expression has a
functional integral representation given by
 \bn  && \<\pi'',g''|\s\E\s|\pi',g'\>
      =\int \<\pi'',g''|{\bf T}\,e^{-i\tint[\s N^a\s\H_a+N\s\H\s]\,d^3\!x\,dt}
  \s|\pi',g'\>\,\D R(N^a,N)\no\\
 &&\hskip1cm=\lim_{\nu\ra\infty}{\o{\cal N}}_\nu\s\int e^{-i\tint[g_{ab}{\dot\pi}^{ab}+N^aH_a+NH]
 \,d^3\!x\,dt}\no\\
  &&\hskip1.5cm\times\exp\{-(1/2\nu)\tint[b(x)^{-1}g_{ab}g_{cd}{\dot\pi}^{bc}{\dot\pi}^{da}+
  b(x)g^{ab}g^{cd}{\dot g}_{bc}{\dot g}_{da}]\,d^3\!x\,dt\}\no\\
  &&\hskip2cm\times[\s \Pi_{x,t}\,\Pi_{a\le b}\,d\pi^{ab}(x,t)\,dg_{ab}(x,t)\s   ]\,
  \D R(N^a,N)\;. \label{f39}\en
Despite the general appearance of
({\ref{f39}), we emphasize once again that this representation has
been based on the affine commutation relations and {\it not} on any
canonical commutation relations.

The expression $\<\pi'',g''|\s\E\s|\pi',g'\>$  denotes the coherent-state
matrix elements of the projection operator $\E$ which projects
onto a subspace of the original Hilbert space on which the quantum
constraints are fulfilled in a regularized fashion. Furthermore, the
expression $\<\pi'',g''|\s\E\s|\pi',g'\>$ is another continuous
 functional that can be used as a reproducing
kernel and thus used directly to generate the reproducing kernel
physical Hilbert space on which the quantum constraints are
fulfilled in a regularized manner.
 Up to a surface term, the phase factor in the
functional integral represents the canonical action for general
relativity, and specifically, $N^a$ and $N$ denote Lagrange
multiplier fields (classically interpreted as the shift and lapse),
while $H_a$ and $H$ denote phase-space symbols (since $\hbar\ne0$)
associated with the quantum diffeomorphism and Hamiltonian
constraint field operators, respectively.

\subsubsection*{The ``diagonal representation''}
It is noteworthy that the connection between the Hamiltonian constraint
operator field ${\cal H}(x)$ and its associated symbol $H(x)$ that is used in
the functional integral (\ref{f39}) is closely related to the
``diagonal representation'' that Sudarshan introduced into quantum optics
\cite{sudar}. In particular,
 \bn   {\cal H}(x)={\o{\cal N}}\int H(x)\;|\pi,g\>\<\pi,g|\,[\s\Pi_x\,\Pi_{a\le b}\,d\pi^{ab}(x)\,dg_{ab}(x)\s]\;. \en
A similar relation connects ${\cal H}_a(x)$ to its symbol $H_a(x)$ for all $a$, $1\le a\le3$.

\subsubsection*{Properties of the regularization}
The $\nu$-dependent factor in the integrand of (\ref{e20}) and (\ref{f39})
formally tends to unity in the limit
$\nu\ra\infty$; but prior to that limit, the given expression
regularizes and essentially gives genuine meaning to the heuristic,
formal functional integral that would otherwise arise if such a
factor were missing altogether \cite{kla2}.  The given form,
 and in particular the
nondynamical, nonvanishing, arbitrarily chosen scalar density
$b(x)$, is very welcome  since this form leads to a reproducing
kernel Hilbert space for
gravity having the needed infinite dimensionality; a seemingly
natural alternative \cite{kla5} using $\sqrt{\det[g_{ab}(x)]}$ in
place of $b(x)$ fails to lead to a reproducing kernel Hilbert space
with the required dimensionality \cite{wat2}. The choice of $b(x)$
determines a specific ultralocal representation for the basic affine
field variables, but this unphysical and temporary representation
disappears entirely after the gravitational constraints are
fully enforced (as soluble examples explicitly demonstrate
\cite{kla3}). The integration over the Lagrange multiplier fields
($N^a$ and $N$) involves a specific measure $R(N^a,N)$,
which is normalized such that $\tint\D
R(N^a,N)=1$. This measure is designed to enforce (a regularized
version of) the quantum constraints$\s$; it is manifestly {\bf
not} chosen to enforce the classical constraints, even in a
regularized form. The consequences of this choice are {\it profound}
in that no (dynamical) gauge fixing is needed, no ghosts are
required, no Dirac brackets are necessary, etc. In short, no
auxiliary structure of any kind is introduced.

\subsubsection*{The gravitational anomaly}
The quantum gravitational constraints, $\H_a(x)$, $1\le a\le3$, and
$\H(x)$, formally satisfy the commutation relations
 \bn &&[\H_a(x),\H_b(y)]=i\s\half\s[\delta_{,a}(x,y)\s\H_b(y)+\delta_{,b}(x,y)\s\H_a(x)]\;,\no\\\
  &&\hskip.15cm[\H_a(x),\H(y)]=i\s\delta_{,a}(x,y)\s\H(y) \;,\\
  &&\hskip.31cm[\H(x),\H(y)]=i\s\quarter\s\delta_{,a}(x,y)\s[\s g^{ab}(x)\s\H_b(x)+\H_b(x)\s g^{ab}(x) \no\\
&&\hskip3.6cm +g^{ab}(y)\s\H_b(y)+\H_b(y)\s g^{ab}(y)\s] \;. \no
\en Following Dirac, we first suppose that
$\H_a(x)\s|\psi\>_{phys}=0$ and $\H(x)\s|\psi\>_{phys}=0$ for all
$x$ and $a$, where $|\psi\>_{phys}$ denotes a vector in the physical
Hilbert space ${\frak H}_{phys}$. However, these conditions are {\it
incompatible} since $[\H_b(x),g^{ab}(x)]\ne0$ and almost surely
$g^{ab}(x)\s|\psi\>_{phys}\not\in{\frak H}_{phys}$, even when
smeared. As noted previously, this means that the quantum
gravitational constraints are partially second class.

\section{Hard-core Picture of Nonrenormalizability}
Nonrenormalizable quantum field theories involve an infinite number of
distinct counterterms when approached by a regularized, renormalized
perturbation analysis. Focusing on scalar field theories, a qualitative
Euclidean functional integral formulation is given by
  \bn  S_\l(h)={\cal N}_\l\int e^{\tint h\s\p\,d^n\!x
-W_o(\p)-\l\s V(\p)}\;{\cal D}\p\;,  \en
where $W_o(\p)\ge0$ denotes the free action and $V(\p)\ge0$ the
interaction term. If $\l=0$, the support of the integral is determined by $W_o(\p)$; when $\l>0$, the support is determined by $W_o(\p)+\l\s V(\p)$.
Formally, as $\l\ra0$, $S_\l(h)\ra S_0(h)$, the functional integral for the free theory. However, it may happen that
   \bn  \lim_{\l\ra0} S_\l(h)=S'_0(h)\not= S_0(h)\;,  \en
where $S'_0(h)$ defines a so-called {\it pseudofree} theory.
Such behavior arises formally
if $V(\p)$ acts as a hard core, projecting out certain fields that are not
restored to the support of the free theory as $\l\ra0$ \cite{kla11}.
In particular, for relativistic $\varphi^4_n$ models, it is known \cite{book},
provided $\p\not\equiv0$, that
  \bn \frac{[\s\tint \p(x)^4\,d^n\!x\s]^{1/2}}{\tint\{\s[\nabla\p(x)]^2
+m^2\s\p(x)^2\s\}\,d^n\!x}\le\frac{4}{3}\;, \en
for $n=3,4$, while for $n\ge5$, no finite upper bound exists. Although
such inequalities are derived for test functions, the bound on
the quotient still applies
to the limit in which a sequence of test functions weakly converges
to a distribution. Such qualitatively different behavior for $n\le4$ and
$n\ge5$ coincides with the division of such models into
renormalizable and nonrenormalizable categories. Based on this fact, it is highly suggestive that nonrenormalizable models have support properties that
are significantly influenced by the hard-core nature of $V(\p)$ relative
to $W_o(\p)$, a property that also accounts for the need of an infinite
set of distinct perturbative counterterms.

It is noteworthy that there exist
highly idealized nonrenormalizable model quantum field theories with
exactly the behavior described; see \cite{book}.
It is our belief that these soluble models strongly suggest that
nonrenormalizable $\varphi^4_n$ models can be understood by the same
mechanism, and that they too can be properly formulated by the incorporation
of a limited number of counterterms distinct from those suggested
by a perturbation treatment. Although technically more complicated, we see
no fundamental obstacle in dealing with quantum gravity on the basis
of an analogous hard-core interpretation. However, that is a problem
for the future.

\section*{Dedication}
I am pleased to dedicate this article to the 75th birthday of George
Sudarshan, and I wish him many more years of good health and productive
research.

\end{document}